# Deviatoric Stress Driven Transient Melting Below the Glass Transition Temperature in Shocked Polymers

Jalen Macatangay, Brenden W. Hamilton, Alejandro Strachan*

School of Materials Engineering and Birck Nanotechnology Center, Purdue University, West Lafayette, IN, 47907, USA

* strachan@purdue.edu

## Abstract

The relaxation of polymers around and below their glass transition temperature is governed by a range of correlated unit processes with a wide range of timescales. The fast deformation rates of shock loading can negate a significant fraction of these processes resulting in the dynamical glass transition in rubbers. In this letter we report the inverse, a transient melting of glassy polymer under shock loading. The large deviatoric stresses near the shock front induce fast transitions in backbone dihedral angles and a stress relaxation characteristic of polymer melts. This is followed by the slower relaxation expected for glasses.

**Introduction**. The thermo-mechanical response of glassy and rubbery polymers is governed by relaxation processes with a broad range of timescales resulting in time-dependent properties. The nature of these relaxation mechanisms and their spatial and temporal correlations across the glass transition temperature ($T_g$) remain poorly understood and are the subject of intensive research. Particularly challenging is the response of polymers to ultra-fast deformation rates such as in high-velocity impact and shock loading, where high strain rates can negate regions of the relaxation spectra. Shock loading takes a material to a state characterized by large, non-hydrostatic stresses and high temperatures in picosecond timescales, which triggers a plethora of response processes. These can include plasticity,[1–4] phase transformations,[5–7] and chemical reactions.[8,9] Interestingly, while many of these processes map to low strain rate processes, exotic phenomena have been observed. Examples include transient melting below the equilibrium melting temperature,[10–13] mechanically-accelerated chemistry,[14–16] and extreme intra-molecular strains.[3,17–19] The triaxiality of the stress state often results in plastic deformation and the underlying mechanisms can be strain-rate dependent.[20] In this paper, molecular dynamics (MD) simulations reveal an unexpectedly fast relaxation during the early stages of shock loading of a glassy polymer. The large triaxial stresses caused by strong shocks induce fast relaxation along the polymers backbone which leads to transient melting and fast relaxation in picosecond timescales.

A well-established, atomic-level understanding of plastic deformation in crystalline systems, described in terms such as dislocation glide,[21–23] grain boundary sliding,[24–26] and shear banding,[20,27,28] undergirds our understanding of material shock responses. Relaxation in glassy polymers involves more complex cooperative segmental relaxation along the backbone and side chains with a wide range of spatial and temporal scales.[29–31] Entangled polymer melts are known to undergo heterogeneous flow through a shear banding process[32,33] and glassy polymers can undergo shear-induced plasticity.[34] Recent experimental breakthroughs are beginning to shed light on the response of rubbery polymers to dynamic impacts that result in ultra-high strain rates (>$10^5$, often ~$10^8$ 1/s).[35] High-velocity impacts in a polybutadiene (PB)-based polyurea (PU) rubber showed a dynamical transition to the glassy state accompanied by high energy absorption.[34] This dynamical glass transition, which results in a mechanical response markedly different from that of lower-$T_g$ PB rubber, is due to the relaxation timescales associated with the PU segments being comparable to the loading rates. Similar trends were observed in high-velocity impact studies on poly(urethane urea) (PUU). These elastomers exhibit restitution coefficients comparable to glassy polycarbonate (PC) even when their storage moduli at ambient conditions differ greatly.[36] Interestingly, no permanent deformation was observed on the PUU surfaces as opposed to PC. These observations are consistent with the general understanding of the relationship between glass transition temperature and the rates of deformation and cooling/heating.[37,38] Fast rates tend to push the glass transition temperature to higher values as described by the classic WLF theory.[39] While the noted response to impacts with macroscale particles results in brittle failure, the observation of hyper-elasticity during microparticle impacts on PUU, with strain rates on the order of $10^8$ 1/s, remains puzzling. This has been attributed to two different segmental relaxation modes in mobile and rigid soft segments.[40] The multiple, local relaxation modes within the segments lead to the unexpected responses for polymers that undergo glass transitions for much slower deformation rates. Our results indicate unexpectedly fast transient relaxation at such fast strain rates even in glasses which can contribute to these observations.

Recent computational work on shocks in PB melts observed *glassification* and identified two different relaxation timescales responsible for shear stress relaxation: a fast mechanism related to intra-molecular relaxations that occur on picosecond timescales and a chain relaxation mechanism that is orders of magnitude slower.[41] However, there is still a considerable knowledge gap concerning the underlying molecular-scale mechanisms and atomic motion that drives these phenomena, and how these phenomena translate to glassy polymers.

**Model**. Our polystyrene system contains 40 chains with 100 monomers each within a simulation cell with 3D periodic boundary conditions. The initial system is built using the continuous configurational bias Monte Carlo approach implemented in Polymer Modeler.[42,43] All-atom MD simulations were conducted with the LAMMPS software package,[44] with the atomic interactions and partial charges described by the Dreiding force field[45] and Gasteiger method,[46] respectively. The model is built at a density of 0.793 g/cm$^3$ and equilibrated at 800 K under isochoric-isothermal conditions (NVT) for 50 ps followed by isothermal-isobaric conditions (NPT) for 500 ps at ambient pressure and 800 K. The system was then cooled in a step-wise manner from 800 K to 300 K, with the temperature decreasing by 10 K every 100 ps under isothermal-isobaric conditions at ambient pressure. Both the glass transition temperature and the room temperature density (1.0 g/cm$^3$) extracted from the simulations are in good agreement with experiments.[47]

**Shock simulations**. From the 300 K equilibrated system, shock loading was conducted using the multiscale shock technique (MSST),[48] which has been previously used to model shock compression events such as phase transformation in silica glass,[49] prebiotic chemistry in planetary impacts,[50] and chemical initiation in energetic materials.[51,52] Compressions were simulated for particle velocities ($u_p$) between 0.3 km/s and 4.1 km/s. Section SM-1 in the Supplementary Material (SM) shows the optimization of parameters associated with MSST. Hugoniot curves in shock velocity vs particle velocity ($U_s$-$u_p$), P-V, and T-$u_p$ planes are included in Section SM-2 of the SM. These results also compare well to experiments.[53]

Non-equilibrium MD (NEMD) shock simulations were also conducted on a slab of PS, created by replicating the original PS cell three times in the shock direction. These simulations were conducted in the reverse ballistic method.[54,55] These simulations are computationally more intensive than MSST but involve an explicity description of the shock front and are used to verify the description of shock transients in MSST. An Eulerian binning was used to compare these spatial results to the temporal evolution of the MSST results.

**Polymer relaxation analysis**. To analyze stress relaxation processes from shock loading, we explored backbone torsional transition events (BTTE), rapid dihedral angle flips between low-energy states within a polymer chain.[56] We define a torsional transition event to occur at time t when the difference between the torsional angles at t − 2 ps and t + 2 ps exceeds 80°. Torsional transition rates were then calculated by normalizing the number of torsional transitions per picosecond by the total dihedrals. We investigated torsional transition rates for transient and steady-state sections of the simulation. The shock transient regime represents the particle velocity-dependent section from shock loading to 2 ps after the local maximum von Mises stress, whereas the shock steady-state represents the last 475 ps of the 500 ps simulation, where steady-state behavior is observed. Figure SM-3 in the SM shows similar results for the last 200 and 400 ps, which indicates the results' insensitivity to the chosen timeframe.

**Results and Discussion.** We characterize the mechanical relaxation of the shocked system by decomposing the stress tensor into its hydrostatic and deviatoric components. Figure 1 shows the time histories of the deviatoric component (also denoted as von Mises stress or J2 invariant). The slow relaxation of deviatoric stress can be contrasted with the hydrostatic component (shown in SM-4) that reaches steady-state within 5 ps for strong shocks with the initial loading completing in approximately 2.5 ps. These timescales are consistent with the timescales observed in our NEMD shock simulations (see SM-4).

For weak shocks, deviatoric stresses relax slowly, as expected of a glassy material. Remarkably, the relaxation of deviatoric stress for shocks with $u_p$ = 1.5 km/s and higher exhibits two distinct regimes. An initial fast relaxation on picosecond timescales is followed by a slower process. The fast initial relaxation is characteristic of materials well above the glass transition temperature,[56] yet the temperature of the shocked systems remains well below the corresponding $T_g$. For example, shocks with $u_p$= 1.5 km/s result in a temperature of 473 K and a pressure of approximately 8 GPa, this can be compared to the zero pressure $T_g$ for our model, 442 K[43] plus the expected increase due to pressure on the order of 1000 K.[41]

To further assess the nature of the fast transient relaxation regime, magenta lines in Figure 1 mark the relaxation dynamics predicted by the Rouse model, that describes unentangled polymer melts. In this case, the time evolution of the deviatoric stress is described by a power law with exponent -0.5.[54] While the short time associated with the transient relaxation precludes us from accurately extracting exponents from the data, we find our MD results for $u_p \geq 1.5$ km/s to be consistent with Rouse relaxation. The cyan lines in Figure 1 show that the post-transient relaxation in strong shocks and the total response for weaker shocks can be described with a single power-law with an exponent of -0.1. Section SM-9 in the SM shows that our data can also be described with a Prony (exponential) series with three terms, which similarly shows fast and slow regimes. The remainder of this letter explores the molecular-scale processes responsible for the observed two-regime relaxation.

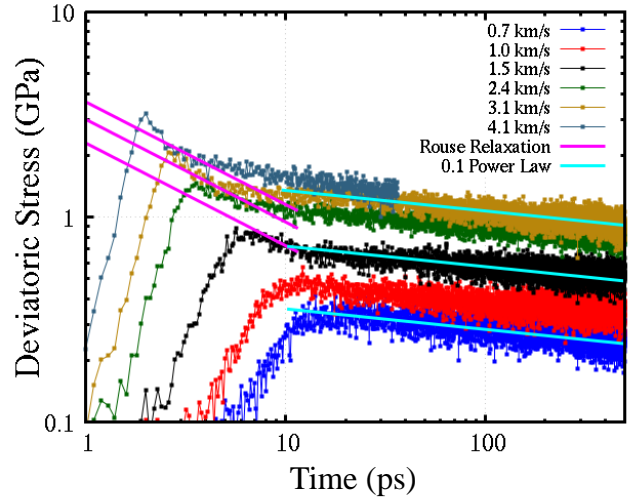

Figure 1: von Mises stress time histories for several $u_p$ shocks. Magenta lines represent the expected decay rates of a Rouse mode relaxation in an unentangled melt for various pre-factors. The cyan lines are similar power law decays with an exponent of -0.1.

MD has been a powerful tool for probing short timescale mechanical effects such as those observed here, as they often occur within the transient state directly behind the wavefront.[57] For shocks in crystalline materials, one proposed mechanism for alleviating large deviatoric stresses is melting below the melt temperature, or 'virtual melting'.[10] A virtual melt (VM) is defined as a short-lived melt of a primary phase that is below its thermodynamic melting temperature for the given hydrostatic pressure.[11] Driven by high deviatoric stresses, VM takes place at temperatures well below the thermodynamic melting temperature to rapidly relax non-hydrostatic stresses. For FCC metals (Cu and Al), VM occurs typically in a nanometer-scale region behind the shock front and is activated by shock strengths on the order of 100 GPa for Cu and roughly 50 GPa for Al.[10] Additionally, VM has been shown to play a role in crystal-crystal and crystal-amorphous phase transformations,[11] as well as interfacial solid-solid phase transformations in the energetic material HMX.[12,13] Local, mechanically-driven melting has also been characterized experimentally as a route for accelerated phase transitions.[58] Motivated by such prior results we hypothesized that the fast transient relaxation following strong shocks could be due to virtual melting. To test this, we characterized the molecular-level relaxation processes during and following shock loading. In particular, we characterized torsional transition events along the polymer backbone, known to control chain relaxation and the glass transition.[59–61] Alzate-Vargas et al. calculated backbone torsional transition rates, the number of torsional transition events per torsion and per unit time, for several polymers across their glass transition temperature. For polystyrene, the threshold torsional transition rate at its glass transition temperature is $10^{-3}$ 1/ps.[56]

Figures 2(a) and (b) show the time evolution of deviatoric stress and BTTE rates for two shock strengths (other cases are shown in Section SM 6 of the SI). Quite interestingly, we find a short (sub-10 ps) burst of BTTE activity following shock loading, followed by a subsequent relaxation to a steady state value. Importantly, high BTTE rates (over an order of magnitude larger than the steady state values) result in fast relaxation of deviatoric stresses. The peak in BTTE transition rates aligns in time with the rapid decrease in deviatoric stress, indicating that such dihedral angle transitions are responsible for stress relaxation. The two-regime stress relaxation behavior is first observed at $u_p$ = 1.5 km/s, which corresponds to the transient BTTE rate surpassing the $T_g$ threshold (the dashed lines in Figure 2(a) and (b) indicate the threshold BTTE rate associated with ambient pressure $T_g$ reported in Ref. [56]).

To confirm the transient effects observed via MSST simulations, we performed a large-scale NEMD simulation. Figure 2(c) shows profiles of hydrostatic and deviatoric stresses and BTTE percentage as a function of position along the shock front at a given time after steady shock propagation has been achieved. The BTTE percentage is the from the raw number of BTT events in the 4 ps window. This NEMD simulation shows the same trends observed in MSST runs. A burst of BTTE activities following the shock front which matches the region of fast relaxation of the deviatoric stress.

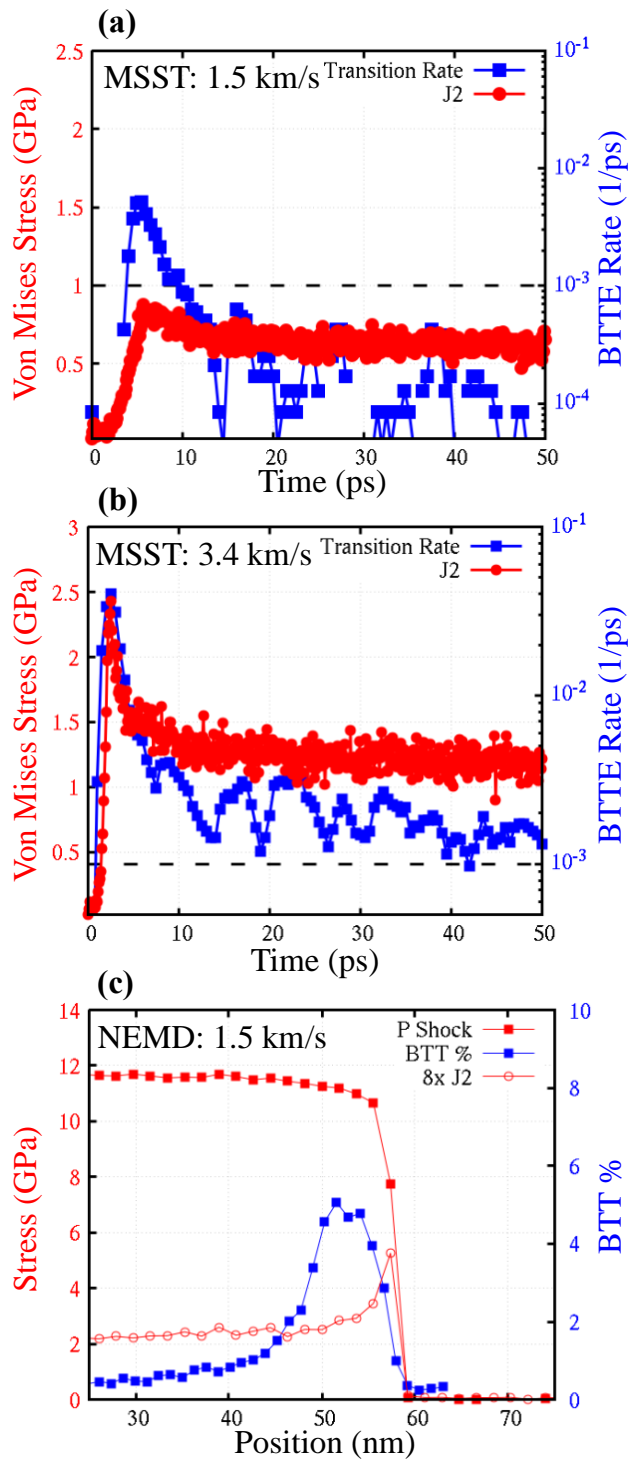

*Figure 2: a & b) Time-averaged torsional transition and von Mises stress at particle velocities of 1.5 and 3.4 km/s using MSST simulations and c) Shock pressure and von Mises stresses and local BTT events (as a percentage of torsions) over space in the shock direction, using an Eulerian binning.*

Figure 3 shows the BTTE rates as a function of $u_p$ for the transient and steady-state regimes. We find that while the initial BTTE rates associated with the shock transient increase monotonously with shock strength (open, red symbols in Figure 3), the steady-state values (blue, closed symbols) begin slightly below the ambient conditions baseline for weak shocks and increase for relatively strong shocks. We attribute the rather constant steady-state BTTEs rates for weak shocks to the interplay between compression and temperature. We find that the transient BTTE rates are significantly greater than the steady-state ones, by roughly an order of magnitude for strong shocks. Quite interestingly, the transient BTTE rates surpass the threshold rate corresponding to the ambient pressure glass transition[56] for particle velocities just below 1.5 km/s, even when the steady-state rates remain well within the glassy region. This supports our hypothesis that transient melting of the polymer is responsible for the fast relaxation events observed in our simulations and prior work.[41,56]

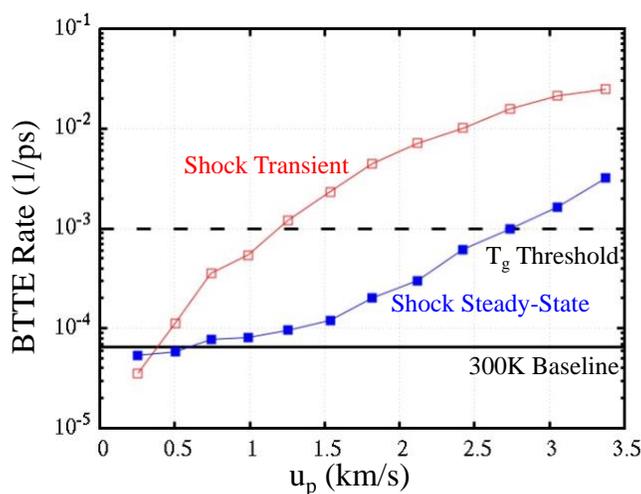

Figure 3: BTTE rates in the transient and steady-state, red and blue symbols respectively

**Summary**. Our results show that in response to the ultrafast development of large deviatoric stress caused by shocks, polystyrene, a glassy polymer, undergoes a short, transient melting which results in rapid stress relaxations. This is mediated by rapid rotations of dihedral angles along the polymer's backbone. As levels of deviatoric stress decrease, so does the rate of chain relaxations, quickly taking the system back to a glassy status where deviatoric stress relaxes on timescales many orders of magnitude longer. This observation is akin to virtual melting reported in crystalline systems[10] but driven by distinct molecular-scale processes. This mechanical activation of BTTEs is perhaps not surprising but, at first sight, may seem to contradict the common trend of polymers being *glassier* for fast strain rates. Our findings indicate that the large deviatoric stresses immediately following shock loading can reduce the timescale associated with relaxation processes by over an order of magnitude causing temporary melting. This burst of relaxation activity quickly decays, and the equilibrium relaxation spectra compared to the loading rate indicate glassier response. The required pressure to initiate virtual melting in polymers shown here is significantly less than what has been previously found for metals. These conditions are relevant for particle impacts related to space exploration,[62,63] hypersonic erosion,[35] needle-free drug delivery,[64] as well as detonation initiation and propagation in energetic materials.[65–67]

## Acknowledgements

This work was primarily supported by the US Office of Naval Research, Multidisciplinary University Research Initiatives (MURI) Program, Contract: N00014-16-1-2557. Program managers: Chad Stoltz and Kenny Lipkowitz. We acknowledge computational resources from nanoHUB and Purdue University through the Network for Computational Nanotechnology.